\documentclass[twocolumn,a4paper,amsmath,amssymb]{revtex4}
\usepackage[urlcolor=blue, hyperindex, colorlinks, bookmarks=true]{hyperref}
\usepackage{graphicx}
\usepackage{times}
\usepackage{amsmath}
\usepackage{braket}
\usepackage{xcolor}
\usepackage{tabularx}

\renewcommand{\eqref}[1]{Eq.~(\ref{#1})}

\newcommand{\figref}[1]{Fig.~\ref{#1}}

\newcommand{\removedD}[1]{{\color{gray}{#1}}}
\renewcommand{\removedD}[1]{{}} 
\newcommand{\new}[1]{{\color{black}{#1}}}

\newcommand{\CEE}[1]{{\color{black}{#1}}}
\newcommand{\corrected}[1]{}

\renewcommand{\eqref}[1]{Eq.~(\ref{#1})}

\newcommand{\appref}[1]{\hyperref[#1]{Appendix~\ref*{#1}}}
\newcommand{\tabref}[1]{\hyperref[#1]{Table~\ref*{#1}}}

\usepackage[english]{babel}
\begin{document}
\title{Electron Spin Resonance at the Level of 10$^4$ Spins\\ Using Low Impedance Superconducting Resonators}
\author{C. Eichler$^1$, A. J. Sigillito$^2$, S. A. Lyon$^2$, J. R. Petta$^1$}
\affiliation{$^1$Department of Physics, Princeton University, Princeton, New Jersey, 08544}
\affiliation{$^2$Department of Electrical Engineering, Princeton University, Princeton, New Jersey, 08544}
\date{\today}
\begin{abstract}
We report on electron spin resonance (ESR) measurements of phosphorus donors localized in a 200 $\mu$m$^2$  area below the inductive wire of a lumped element superconducting resonator. By combining quantum limited parametric amplification with a low impedance microwave resonator design we are able to detect around  \CEE{$2\times10^4$} spins with a signal-to-noise ratio (SNR) of 1 in a single shot.  The 150 Hz coupling strength between the resonator field and individual spins is significantly larger than the 1 -- 10 Hz coupling rates obtained with typical coplanar waveguide resonator designs. Due to the larger coupling rate, we find that spin relaxation is dominated by radiative decay into the resonator and dependent upon the spin-resonator detuning, as predicted by Purcell.
\end{abstract}
\maketitle
Electron and nuclear spin magnetic resonance are widely used to characterize a diverse set of paramagnetic materials in biology, chemistry, and physics \cite{Chizhik2014}. They also play an important role in the control and readout of spin qubits as highly coherent carriers of quantum information \cite{Tyryshkin2011,Maurer2012}. Improving the sensitivity of spin resonance is an outstanding goal, which has triggered research on a variety of measurement schemes such as optical \cite{Wrachtrup1993,LeSage2013}, electrical \cite{Elzerman2004,Hoehne2010,Morello2010}, and mechanical \cite{Barfuss2015} detection.

Important progress has also been made by inductively coupling spins to superconducting resonators through the magnetic dipole interaction \cite{Kubo2010} and by adopting ideas and techniques from circuit quantum electrodynamics \cite{Wallraff2004,Blais2004}. Strong collective spin coupling with superconducting resonators and qubits has been  demonstrated in various materials such as nitrogen vacancy and P1 centers in diamond \cite{Kubo2010,Schuster2010a,Zhu2011}, rare earth ions \cite{Probst2013}, ferromagnets  \cite{Tabuchi2015}, and dopants in silicon \cite{Zollitsch2015}. The number of spins involved in most of these experiments is typically $10^{10}$ to $10^{13}$, far from the single spin limit. In an effort to reduce the number of spins and improve ESR sensitivity, remarkable achievements have recently been made by coupling bismuth donors in silicon to resonators with quality factors exceeding 10$^{5}$ \cite{Bienfait2016,Bienfait2016a}.

In this Letter, we demonstrate a complementary approach to improve ESR sensitivity that is based on the enhancement of the single spin coupling strength to the resonator field by using lumped element resonators \cite{Probst2013,Bienfait2016} with reduced characteristic impedance. \new{Increasing the coupling strength is particularly helpful when only moderate quality factors are achievable. This is often the case in the presence of large magnetic fields that are required to achieve spin transitions in the microwave frequency range for the majority of spin species.}
The measurements presented here are performed in magnetic fields $B_0$ $\approx$ 180 mT with phosphorus donors in isotopically purified $^{28}$Si, which are representative of the class of spin systems with a $g$-factor close to 2. By integrating a Josephson parametric amplifier (JPA) into the detection chain we demonstrate the detection of about $2\times 10^4$ electron spins with a SNR of 1, which
exceeds previously reported sensitivities in \new{phosphorus doped silicon} by more than two orders of magnitude \cite{Sigillito2014}. We also measure the dependence of the spin lifetime $T_1$ on the spin-resonator detuning and find it to be limited by Purcell decay \cite{Purcell1946}, in accordance with previous observations \cite{Bienfait2016a,Putz2014}.

\begin{figure*}[t]
\includegraphics[scale=1]{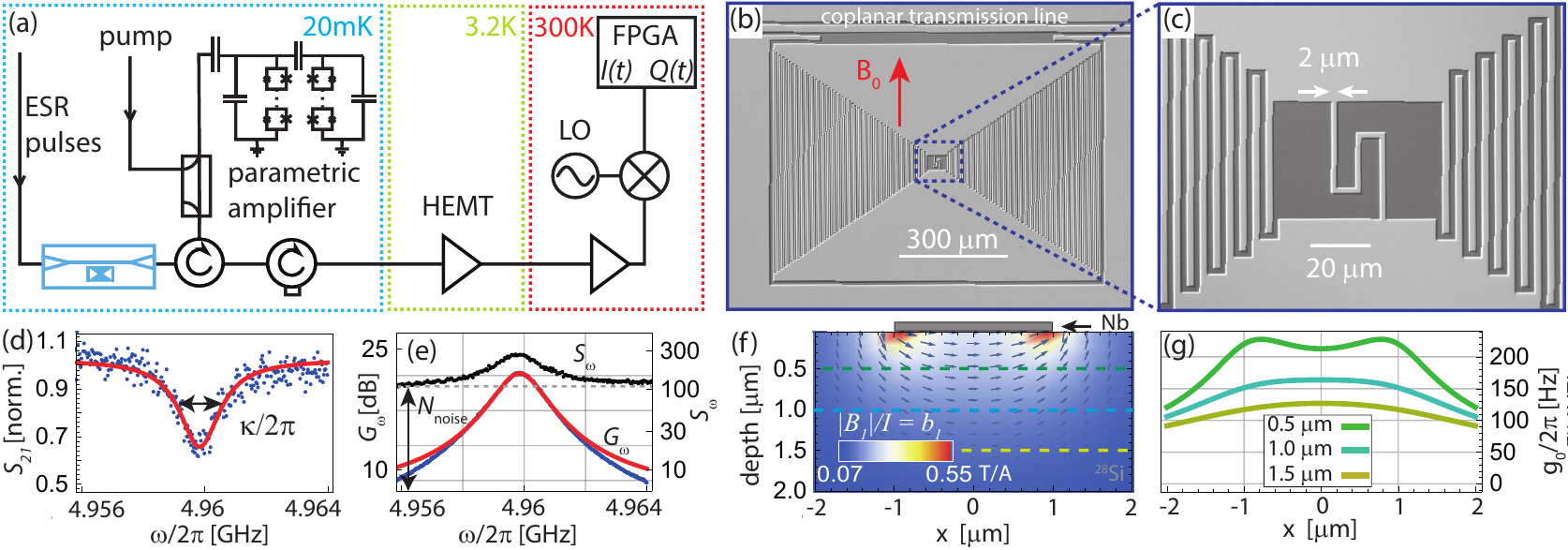}
\caption{ (a) Schematic of the experimental setup. The output field of the resonator is amplified by a JPA at base temperature, further amplified by a high-electron mobility transistor (HEMT) amplifier at $3.2\,$K and demodulated with an IQ mixer at room temperature to record and process the quadratures $I(t)$ and $Q(t)$ of the signal with FPGA electronics. (b) Optical micrograph of a resonator with low characteristic impedance $Z=\sqrt{L/C} \approx$ 8 $\Omega$ that is side-coupled to a coplanar transmission line for control and readout. Coupling between the spins and the resonator field occurs below the inductive wire shown with higher magnification in (c). (d)  Measurement (blue dots) and Lorentizan fit (red line) of the resonator transmission. (e) Measurement (blue dots) and Lorentzian fit (red line) of the JPA gain $G_\omega$. Measured noise power spectral density $S_\omega$ (black dots) in units of \emph{photons per Hz per second}. (f) Simulated magnetic field distribution per current $b_1 = B_1/I$ in the phosphorus doped region underneath the wire. (g) Simulated coupling strengths $g_0$ at three different depths below the wire, as indicated by the dashed lines in panel (f).}
\label{fig:1}
\end{figure*}
Our experiments take place using the cryogenic setup depicted in \figref{fig:1}(a). Measurements are performed using planar superconducting microwave resonators with low characteristic impedance $Z$ $\sim$ 8 $\Omega$ [see \figref{fig:1}(b)]. The triangular shaped regions form interdigitated finger capacitors that shunt an inductive wire of length $l\approx 100\, \mu$m and width $w \approx 2\,\mu$m, as shown in \figref{fig:1}(c). The oscillating magnetic field associated with photons in the resonator is thus strongly confined in the region below the inductive wire.  The resonator is fabricated from a niobium thin film sputtered on top of a 2$\,\mu$m thick layer of epitaxially grown $^{28}$Si uniformly doped during growth with $5 \times 10^{15}$ phosphorus donor atoms per cubic centimeter \new{and with a residual $^{29}$Si content of 800 ppm}. The coupling strength $g_0$ between the resonator and individual spins is proportional to the oscillating magnetic field per unit current $b_1\equiv B_1/I$, of which we show the simulated cross-sectional distribution in the spin doped area below the inductive wire [see \figref{fig:1}(f)]. The non-uniform distribution of $B_1$ together with the resonator's estimated characteristic impedance  $Z=\sqrt{L/C} \approx 8\,\Omega$ \cite{Eichler2016SM}
 results in a position-dependent spin-resonator coupling strength \new{$g_0 = b_1 g \mu_B \omega_{\rm res}/\sqrt{8 \hbar Z}$} [see \figref{fig:1}(g)]. Here $\omega_{\rm res}$ is the resonator frequency, $g\approx 2$ is the electron $g$-factor in silicon, and $\mu_B$ is the Bohr magneton.

The resonator is side-coupled to a coplanar transmission line that is used for spin control and readout. We apply pulsed and continuous microwave fields through a highly attenuated input line and use a linear detection chain to detect the two field quadratures $I(t)$ and $Q(t)$ of the radiation emitted from the sample, see \figref{fig:1}(a). A static in-plane magnetic field $B_0$ is applied by a superconducting magnet.  With $B_0 \approx 180\,$mT and low probe powers $P \approx -130\,$dBm we measure a resonance frequency close to $\omega_{\rm res}/2\pi = 4.96\,$GHz. The resonator linewidth $\kappa/2\pi = 1.6\,$MHz corresponds to a loaded quality factor of 3100, see \figref{fig:1}(d). The reduction in the transmission at resonance allows us to distinguish between the internal loss rate $\kappa_{i}/2\pi = 1.3\,$MHz and the loss rate due to coupling to the transmission line $\kappa_{\rm ext}/2\pi = 0.3\,$MHz \cite{Gardiner1985}. \new{The internal loss rate is most likely limited by magnetic field-induced losses due to a finite out-of-plane component and by dielectric losses in the substrate. Even at a constant internal loss rate the sensitivity could be enhanced further by designing the external loss rate to be equal and by using asymmetrically coupled drive and readout ports.}

An essential aspect of our experiments is the incorporation of a JPA in the detection chain \cite{Bienfait2016}.  By pushing the linear detection efficiency to its ultimate quantum limit, these amplifiers have recently triggered a paradigm shift in microwave frequency measurements at mK temperatures \cite{Yurke2006,Castellanos2008}, with applications  in superconducting \cite{Vijay2011,Riste2012,Eichler2012b}, semiconducting \cite{Stehlik2015}, and electromechanical systems \cite{Teufel2011a}. The JPA used in this experiment is based on coupled nonlinear resonators and amplifies incoming signals independent of their phase \cite{Eichler2014a}. The JPA gain, shown as blue dots in \figref{fig:1}(e), exhibits a Lorentzian shaped frequency dependence (red line) and is tuned such that the maximum power gain of $G=22\,$dB coincides with the  resonator frequency. To characterize the improvement in the detection efficiency enabled by the JPA we measure the noise power spectral density at the end of the detection chain [black dots in \figref{fig:1}(e)]. On top of the frequency independent noise offset originating from the secondary amplifier stages, the parametric amplifier causes a noise rise proportional to its gain \cite{Eichler2016SM}. This noise contribution has been demonstrated to arise from  vacuum fluctuations at the input of the amplifier \cite{Eichler2014a}. 
By comparing the background noise offset $N_{\rm noise}$ with the parametrically amplified vacuum noise contribution, we approximate the detection efficiency of the amplification chain to be $\eta_{\rm amp} \approx 1/(1+G/N_{\rm noise}) \approx 59\%$ at $\omega_{\rm res}$. Based on room temperature measurements of cable attenuations, we estimate a transmission efficiency $\eta_{\rm loss} \approx 50\%$ between the resonator output and parametric amplifier input. This leads to a combined detection efficiency of $\eta = \eta_{\rm amp} \eta_{\rm loss}\ \approx 29\%$ for the entire detection chain, which is about 60 times larger than the detection efficiency with the JPA turned off.

We probe the phosphorus donors by employing spin echo measurement techniques. The spins are initially spin polarized due to the large Zeeman field. An initial $\pi/2$-pulse is then applied at frequency $\omega_{\rm res}$ to rotate the spins to the equator of the Bloch sphere, where they precess at the Larmor frequency for a time $\tau$. A $\pi$-pulse is applied to reverse the phase accumulation from the preceding free evolution interval. The resulting build-up of a large total spin coherence $S^-(t) = \sum_i{\sigma_i^-(t)}$, where $\sigma_i^-$ denote spin lowering operators of individual spins,  is known as the spin echo \cite{Hahn1950}. In the limit where the effective echo duration $2T_2^*$ is large compared to the resonator decay time $1/\kappa$, the resonator field $a$ follows the total spin coherence $S^{-}$ quasi-instantaneously, so that the two are approximately proportional $a(t) \approx 2  S^{-}(t) \bar{g}/\kappa$ with $\bar{g}$ being the average coupling strength between the participating spins and the resonator \cite{Bienfait2016}.

\begin{figure}[t!]
\includegraphics[scale=.90]{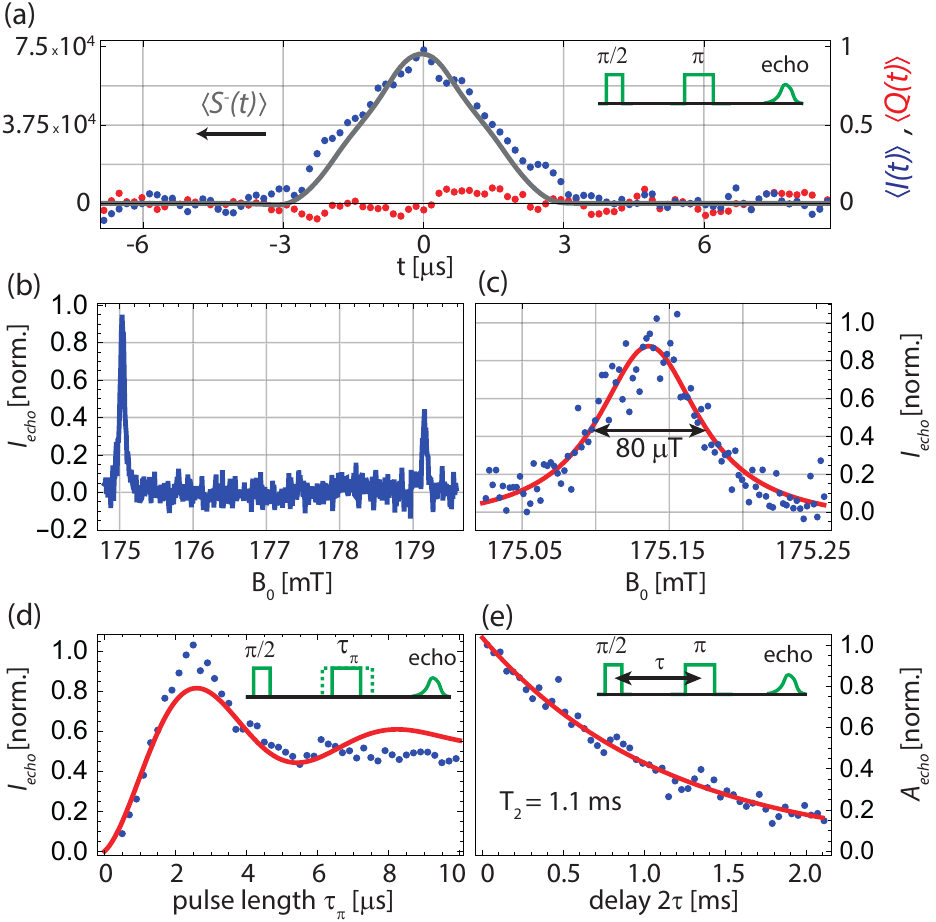}
\caption{ \new{(a) Measured quadrature signal (dots) together with a numerical simulation of $\langle S^{-}(t)\rangle$ (gray line). The echo duration is about $2T_2^*\approx 4\,\mu$s. The data have been taken with the JPA turned off and by averaging over a thousand single shot time-traces. (b) Integrated echo signal $I_{\rm echo}$ as a function of magnetic field $B_0$. The asymmetry in the echo response for the two hyperfine peaks is predominantly explained by a change in the resonator frequency when sweeping $B_0$.} (c) High resolution measurement of $I_{\rm echo}$ around the low field electron spin transition (blue dots) and a Lorentzian fit (red line). (d)  $I_{\rm echo}$ as a function of $\pi$-pulse length $\tau_{\pi}$. The optimal $\tau_{\pi}$  = $2.4\,\mu$s. (e) $A_{\rm echo}$ as a function of delay time $\tau$ between the $\pi/2$ and the $\pi$-pulse . Data are fit by an exponential decay (red line), with a best fit of $T_2$ = 1.1 ms.}
\label{fig:2}
\end{figure}

We measure the spin echo by detecting the quadratures $I(t)$ and $Q(t)$ of the resonator output field as shown in \figref{fig:2}(a). To maximize the SNR we apply a mode-matched filter function $f(t)= \langle I(t)\rangle/(\int{{\rm d}t  \langle I(t)\rangle^2})^{1/2}$ to the time-resolved data and extract the integrated echo signal $I_{\rm echo} = \int {\rm d}t {f(t)} I(t)$ for each  measurement trace recorded in a single shot fashion with a repetition rate of $0.1\,$Hz. Here, $\langle ... \rangle$ corresponds to an ensemble average over multiple single shot time-traces.
The average $\langle I(t) \rangle$ data shown in the inset
are in very good agreement with the time-dependent total spin coherence $\langle S^{-}(t)\rangle$ (gray line), which we have simulated by taking into account the inhomogeneous distribution of  both the spin transition frequencies and the coupling strengths as illustrated in \figref{fig:1}(f) and (g) \cite{Eichler2016SM}. For the chosen pulse parameters and donor density the simulated spin echo signal at its maximum is about $\langle S^{-}(t=0)\rangle\approx 75000$ as shown in \figref{fig:2}(a). This corresponds to an average of $N\approx 150000$ spins contributing to the echo.  While inhomogeneities in the transition frequencies predominantly arise from the non-uniform field $B_0$, the inhomogeneous coupling strength is caused by the position dependent field $B_1$. The latter could be improved with further advances in the resonator design \cite{Mohebbi2014}.

We first  perform magnetic field spectroscopy by measuring the integrated echo signal $I_{\rm echo} $ as a function of $B_0$. We observe two spin resonance transitions that are detuned by $4.2\,$mT, which correspond to the two nuclear spin states of the phosphorus donor atom. The hyperfine coupling rate $A/\hbar = 117.6\,{\rm MHz}\;\widehat{=}\;4.2\,{\rm mT}$ is in excellent agreement with the measured splitting. A higher resolution scan over the lower field transition [\figref{fig:2}(c)] reveals a Lorentzian lineshape with an inhomogeneous linewidth of about 80 $\mu$T, \new{which is supposedly limited by inhomogeneity in the $B_0$ field and corresponds to a dephasing rate $\Gamma/2\pi$ = 2.2 MHz}. The Zeeman energy of the spin ensemble is tuned into resonance with $\omega_{\rm res}$ by setting $B_0 = 175.14\,$mT.

The optimal $\pi$-pulse length is determined by measuring $I_{\rm echo}$ as function of $\tau_{\pi}$. These data are shown in  \figref{fig:2}(d) for a pulse power \CEE{$P_{\rm pulse} \approx -77\,$dBm}. The resulting Rabi oscillations plateau after about half a period due to the inhomogeneous distribution of coupling strengths in the device \cite{Sigillito2014}. The maximum echo signal is obtained for $\tau_{\pi}$ = $2.4\,\mu$s, which we use for the following data sets. We measure the spin coherence time $T_2$ by varying the time delay $\tau$ between the $\pi/2$-pulse and the refocusing $\pi$-pulse. For this particular experiment we average the amplitude $A_{\rm echo}=\int {\rm d}t f(t)\sqrt{\langle I^2(t) + Q^2(t)\rangle}$ in order to compensate for decoherence caused by low frequency magnetic field noise. The resulting data points exhibit an exponential decay with a best fit of $T_2$ = 1.1 ms, which is most likely limited by dipolar interactions between neighboring spins \new{leading to instantaneous diffusion} \cite{Tyryshkin2011}.

\begin{figure}[t!]
\includegraphics[scale=0.9]{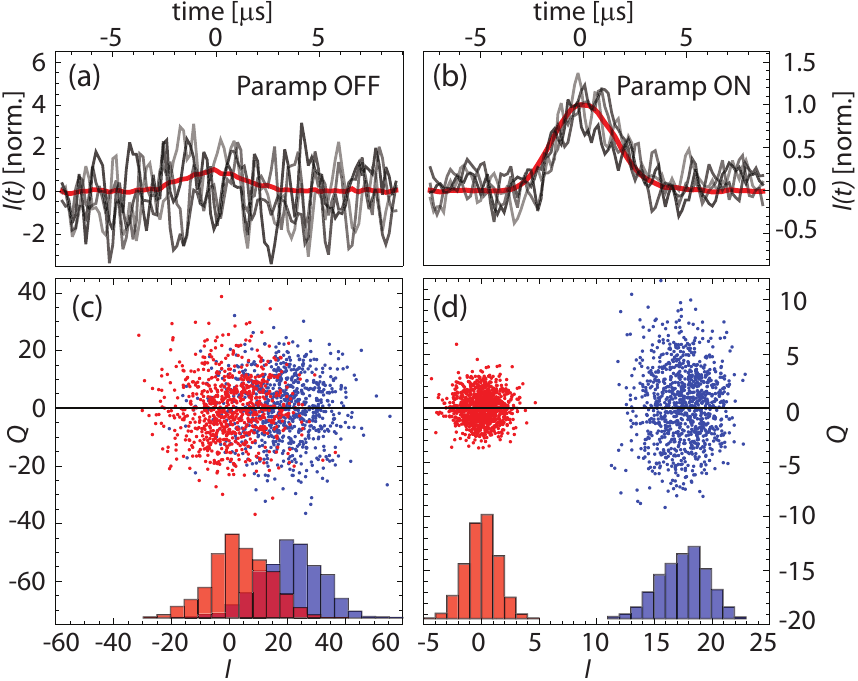}
\caption{Characteristic single shot time-traces of the echo signal (black lines) and an average over 1000 single shot traces (red line) with the JPA turned off (a) and on (b). (c) - (d) Individual measurement results for the filtered echo signal in the $IQ$ plane (blue dots) in comparison with the detection noise (red dots). From the distribution along the $I$ axis we extract a SNR of $\bar{I}/\Delta I  = 10$ with the JPA turned on.}
\label{fig:3}
\end{figure}

In order to study the ESR sensitivity and the improvements enabled by the use of a JPA, we repeatedly measure single shot time-traces of the echo response. From the characteristic time-traces, shown as black lines in Figs.~\ref{fig:3}(a) and (b), we immediately see the significant improvements in SNR with the JPA. With the JPA turned off the echo signal is only resolved after averaging (red line). The echo signal, however, clearly exceeds the noise level in single measurement shots when the JPA is turned on.

We further quantify the SNR by applying the optimal filter function to the time-resolved data resulting in one pair of quadratures $\{I,Q\}$ for each time-trace. We plot 500 such pairs in the $IQ$ plane (red points) in comparison with  measurements of the background noise (blue points), which is recorded $40\,\mu{\rm s}\gg T_2^*$ after each spin echo [Figs.\ \ref{fig:3}(c) and (d)]. The phase reference is chosen such that the echo signal is entirely in the $I$ quadrature. The $I$ and $Q$ axes are scaled to correspond to the real and imaginary part of the detected mode referenced to the output of the cavity. From the standard deviation $\Delta I=11.1\,(1.7)$ and the mean $\bar{I}=18.6\,(17.5)$ in the $I$ quadrature we extract a SNR of $\bar{I}/\Delta I = 1.7\,(10)$ with the JPA turned off (on). The increase in the variations in the $Q$ quadrature for the echo signal visible in \figref{fig:3}(d) is explained by slow phase drifts during the time of data acquisition which is about 2 hours for the data points shown. In addition to the increased variance in the $Q$ quadrature, we also find that the $I$ quadrature of the echo signal exhibits larger variations than the bare detection noise hinting at variations in the number of spins participating in the echo sequence from measurement to measurement.

Based on the SNR measurements and the simulated number of spins $N=150000$ that are excited during the spin echo, we can estimate the ultimate spin sensitivity of the low impedance resonator. With the JPA on we find an ESR sensitivity of approximately $1/N_{\rm min}\approx 1/15000$. We can compare this value to the theoretically estimated value of $1/N_{\rm min} \approx g_0 \sqrt{\eta_{\rm} T_{2}^* \kappa_{\rm ext}/\kappa^2}$ \cite{Bienfait2016}, which for our sample parameters is $\sim 1/10000$ and in reasonable agreement with the measured sensitivity.
\begin{figure}
\includegraphics[scale=0.9]{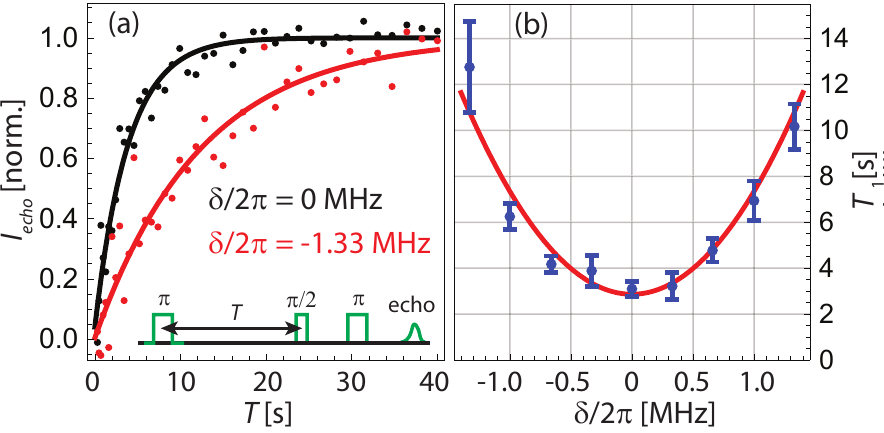}
\caption{$T_1$ as a function of $\delta = E_z/\hbar - \omega_{\rm res}$. (a) Two characteristic data sets from which the $T_1$ times are extracted. Inset: Population recovery pulse sequence. (b) Comparison between the measured lifetimes (blue dots) and the Purcell prediction (red line) assuming $g_0/2\pi =150\,$Hz and the independently measured $\kappa/2\pi = 1.6\,$MHz. Each data point results from the average of two values independently fitted to data sets as shown in panel (a). Error bars indicate the standard error extracted from the exponential fit.}
\label{fig:4}
\end{figure}

As demonstrated above, the increased ESR sensitivity that we achieve is predominantly due to the enhanced single spin coupling strength $g_0$. As predicted by Purcell, the spin--resonator coupling induces an additional radiative decay channel for the excited spins with a Purcell enchanced decay rate $\Gamma_P = 4 \kappa g_0^2/{(\kappa^2 + 4 \delta^2)}$, where $\delta = E_z/\hbar - \omega_{\rm res} $ is the spin-resonator detuning \cite{Butler2011}. As has recently been demonstrated \cite{Bienfait2016a}, this decay channel can significantly dominate over intrinsic $T_1$ decay mechanisms, which were shown many decades ago to yield lifetimes as long as hours \cite{Feher1959}. Acceleration of the spin relaxation rate may be useful for the initialization of spin states in quantum computing protocols.

To investigate the possibility of a Purcell enhancement we have measured the spin lifetime in a population recovery experiment for various detunings $\delta$. \new{The detuning is kept constant during each experiment by setting the magnetic field $B_0$ accordingly}. The spin rotation pulses are applied resonantly with the spin transition and with powers adjusted to the detunings in order to account for pulse filtering by the resonator. As depicted in the inset of \figref{fig:4}(a) an initial $\pi$-pulse inverts the spin population. After a variable time $T$ the recovery of the spin population is probed with a standard Hahn echo sequence to determine what fraction of the spins have relaxed back to the ground state.  As expected, the spin population is well fitted by an exponential function (solid lines) from which we extract the spin $T_1$. At zero detuning $\delta = 0$ the spin $T_1$ is about 3 seconds. However, the spin $T_1$ at $\delta/2\pi = -1.33\,$MHz is measured to be about four times larger. We have repeated such population recovery measurements for various values of $\delta$, see \figref{fig:4}(b). A comparison with the expected Purcell lifetime $1/\Gamma_P$ (red solid line) shows good qualitative agreement with the data, which suggests that the $T_1$ is limited by Purcell decay into the resonator. Our measurements are in agreement with previous measurements of bismuth donors in silicon \cite{Bienfait2016a}.

In conclusion, we have demonstrated significant improvements in the ESR detection sensitivity of phosphorus donors in silicon using lithographically defined resonators. This has been achieved by enhancing the single spin coupling strength with an optimized resonator design and by employing state-of-the-art parametric amplification. Further improvements in the sensitivity could be achieved by reducing the width of the inductive wire and by increasing the quality factor of the resonator. The increased coupling rate of the low impedance resonator design allows for significant tuning of the spin relaxation rate by taking advantage of the Purcell effect.

 We acknowledge Thomas Schenkel for providing the isotopically enriched $^{28}$Si sample. Supported by the Gordon and Betty Moore Foundation's EPiQS Initiative through Grant GBMF4535, with partial support from the National Science Foundation (DMR-1409556 and DMR-1420541). Devices were fabricated in the Princeton University Quantum Device Nanofabrication Laboratory.
\\

\bibliographystyle{apsrev4-1}
%

\end{document}